\documentclass[12pt]{article}
\usepackage{epsf,amsmath}

\begin{document}

\title{CHARM CONTENT OF A PROTON IN COLLINEAR PARTON MODEL AND
IN $K_T-$FACTORIZATION APPROACH
\thanks{This work is
supported by the RFBR under Grant 02-02-16253}}
\author{V.A.Saleev and D.V.Vasin \\
Samara State University \\
Samara 443011, Russia\\
E-mail: saleev@ssu.samara.ru and vasin@ssu.samara.ru}

\maketitle
\begin{abstract}
\noindent  It is shown that the difference between the c-quark
proton SF's calculated in the $k_T$-factorization approach using
different unintegrated gluon distribution functions is the same
order as the difference between results obtained in the parton
model and in the $k_T$-factorization approach.

\end{abstract}

\section{Introduction} The result of a study for the internal structure
of a proton in the process of the lepton deep inelastic
scattering (DIS) can be presented in terms of a proton structure
function (SF) $F_2^p(x_B,Q^2)$ as a function of $Q^2=-q^2$ and
$x_B=Q^2/2(pq)$, where $q$ is the exchange photon 4-momentum and
$p$ is the proton 4-momentum. In a process of the charmed quark
leptoproduction the charmed content of the proton structure
function $F_{2c}^p(x_B,Q^2)$ is probed. The recent relevant
measurings by the H1 \cite{H1} and the ZEUS \cite{ZEUS}
Collaborations at the HERA ep-collider include the following
kinematic region: $1.8<Q^2<130$ GeV$^2$ and $5\cdot
10^{-5}<x_B<2\cdot 10^{-2}$.

The charmed quark SF has been studied in the framework of DGLAP
\cite{DGLAP} and BFKL \cite{BFKL} dynamics. Usually, the c-quark
SF $F_{2c}^p(x_B,Q^2)$ is calculated via the amplitude which
described by the quark box diagrams. This type of a calculation
for the $F_{2c}^p(x_B,Q^2)$ is presented in the talk by A.Kotikov
\cite{Kot}.

Here we use another method which is based on a direct calculation
of the total $c\bar c-$production cross section in the electron
DIS. In a such way, we have obtained the c-quark distribution
function $C_p(x_B,Q^2)$ which is connected with the c-quark SF as
follows:
\begin{equation}
F_{2c}^p(x_B,Q^2)=2e_c^2x_BC_p(x_B,Q^2).
\end{equation}

\section{Electroproduction cross section}

In the framework of the parton model and the one photon exchange
approximation the charmed quark production cross section in the
electron DIS can be presented as a convolution of the c-quark
proton distribution function and the electron -- c-quark partonic
cross section:
\begin{equation}
d\sigma(ep\to ecX)=\int dx_BC_p(x_B,Q^2)d\hat \sigma (ec\to ec).
\end{equation}

The doubly differential cross section can be presented as follows:
\begin{equation}
\frac{d\sigma}{dx_BdQ^2}(ep\to ecX)=C_p(x_B,Q^2)\frac{\overline
{|M(ec\to ec)|^2}}{16\pi(x_Bs)^2},
\end{equation}
where $s=(p_e+p)^2$, $p$ is the proton 4-momentum, $p_e$ is the
electron 4-momentum. The squared amplitude of an elastic
$ec-$scattering has the following form:
\begin{equation}
\overline {|M(ec\to
ec)|^2}=2\frac{e^4e_c^2}{Q^4}(x_Bs)^2\left(y^2-2y+2-\frac{2m_c^2y^2}{Q^2}\right),
\end{equation}
where $y=Q^2/(x_Bs)$. From (1), (3) and (4) we can obtain the
master formula
\begin{equation}
F_{2c}^p(x_B,Q^2)={x_BQ^4\displaystyle\frac{d\sigma}{dx_BdQ^2}(ep\to
ecX )}/\left({\pi\alpha^2
(y^2-2y+2-\displaystyle\frac{2m_c^2y^2}{Q^2})}\right).
\end{equation}
%
%
At the high energy the dominant mechanism of the c-quark
electroproduction on a proton is the photon-gluon fusion. In the
leading order approximation for the QCD running constant
$\alpha_s$ the relevant subprocess is $e+g\to e+ c +\bar c$.

In the conventional collinear parton model it is suggested that
hadronic cross section, in our case $\sigma (ep\to ecX, s)$, and
the relevant partonic cross section $\hat \sigma(eg\to ec\bar
c,\hat s)$ are connected as follows:
\begin{equation}
\sigma^{PM}(ep\to ecX,s)=\int dx  G(x,\mu^2)\hat \sigma(eg\to
ec\bar c,\hat s),
\end{equation}
where $\hat s=xs$, $G(x,\mu^2)$ is the collinear gluon
distribution function in a proton, $x$ is the gluon fraction of a
proton momentum, $\mu^2$ is the typical scale of a hard process.
The $\mu^2$ evolution of the gluon distribution $G(x,\mu^2)$ is
described by DGLAP evolution equation \cite{DGLAP}. In the
$k_T$-factorization approach hadronic and partonic cross sections
are related by the following condition \cite{Lim}:
\begin{eqnarray}
\sigma^{KT}(ep \to ecX)=\int \frac{dx}{x}\int d{\vec k_{T}^2}\int
\frac{d\phi}{2\pi}\Phi(x,\vec k_{T}^2,\mu^2) \hat \sigma(eg^*\to
ec\bar c, \hat s)
\end{eqnarray}
where $\hat \sigma(eg^*\to ec\bar c, \hat s)$ is the c-quark
production cross section on the off mass-shell ("reggeized")
gluon, $k^2=-\vec k_{T}^2$, $\hat s=xs-\vec k_{T}^2$, $\phi$ is
the azimuthal angle in the transverse $XOY$ plane between vectors
$\vec k_{T}$ and the fixed $OX$ axis ($\vec p_e \mbox{ and } \vec
p_e' \in XOZ$).

The unintegrated gluon distribution function $\Phi(x,\vec
k_{T}^2,\mu^2)$ satisfies the \\ BFKL evolution equation [4]. At
the $x\ll 1$ the off mass-shell gluon has dominant longitudinal
polarization along a proton momentum and the gluon polarization
four-vector is written as follows \cite{Lim}
$\varepsilon^{\mu}(k)={k^{\mu}_T}/{|\vec k_T|}$.

 Our calculation
in the parton model was done using the GRV~\cite{GRVLO} and the
CTEQ5L~\cite{CTEQ5L} parameterizations for a collinear gluon
distribution function $G(x,\mu^2)$. In case of the
$k_T$-factorization approach we use the following
parameterizations for an unintegrated gluon distribution function
$\Phi(x,\vec k_{T}^2,\mu^2)$: JB by Bluemlein \cite{JB}, JS by
Jung and Salam \cite{JS}, KMR by Kimber, Martin and Ryskin
\cite{KMR}. We compared these parameterizations directly in our
recent paper \cite{VS}.

Finally, in the $k_T-$factorization formalism the doubly
differential cross section for the process $ep\to ecX$ can be
written as follows:
\begin{equation}
\frac{d\sigma^{KT}}{dx_BdQ^2}=\frac{y}{x_B}\int dp_{cT} d\phi_c
d\eta_c d\vec k_T^2\frac{d\phi}{2\pi}
\frac{p_cp_{cT}}{E_c}\frac{\overline{|M(eg^*\to ec\bar
c)|^2}}{256\pi^4(y-a_1)(xs)^2}\Phi(x,\vec k_T^2,\mu^2),
\end{equation}
where $p_c=(E_c,\vec p_c)$ is the c-quark 4-momentum, $\eta_c$ is
the c-quark pseudorapidity, $\phi_c$ is the azimuthal angle
between OX axis and vector $\vec p_{cT}$, $a_1=2(pp_c)/s$,
$b_1=2(p_ep_c)/s$ and
\begin{equation}
 x=({\vec k_T^2+Q^2+yb_1s+2(\vec q_T\vec k_T)-2(\vec
p_{cT}\vec k_T)-2(\vec q_T\vec p_{cT})})/({(y-a_1)s}).
\end{equation}
We use the following approximations for gluon 4-momentum
$k^{\mu}=xp^{\mu}+k_{T}^{\mu}$, where $k_T^{\mu}=(0,\vec k_T,0)$.

In the parton model one has $\vec k_T=0$ and
\begin{eqnarray}
\frac{d\sigma^{PM}}{dx_BdQ^2}=\frac{y}{x_B}\int dp_{cT} d\phi_c
d\eta_c
\left(\frac{p_cp_{cT}}{E_c}\right)\frac{\overline{|M(eg\to ec\bar
c)|^2}}{256\pi^4(y-a_1)(xs)^2}xG(x,\mu^2),
\end{eqnarray}
where
\begin{equation}
 x=({Q^2+yb_1s-2(\vec q_T\vec p_{cT})})/({(y-a_1)s}).
\end{equation}
The obtained results (Fig.~1) demonstrate agreement between our
predictions and the recent data for the $F_{2c}^p(x_B,Q^2)$ from
HERA \cite{ZEUS}. However, we see that the difference between the
c-quark proton SF's calculated in the $k_T$-factorization approach
using different unintegrated gluon distribution functions is the
same order as than the difference between results obtained in the
parton model and in the $k_T$-factorization approach.

The authors would like to thank B.~Kniehl, A.~Kotikov and H.~Jung
for discussion of the obtained results,  L.~Lipatov and V.~Kim
for kind hospitality during Workshop DIS-2003.

\newpage

\begin{figure}[t]

\begin{center}
\includegraphics{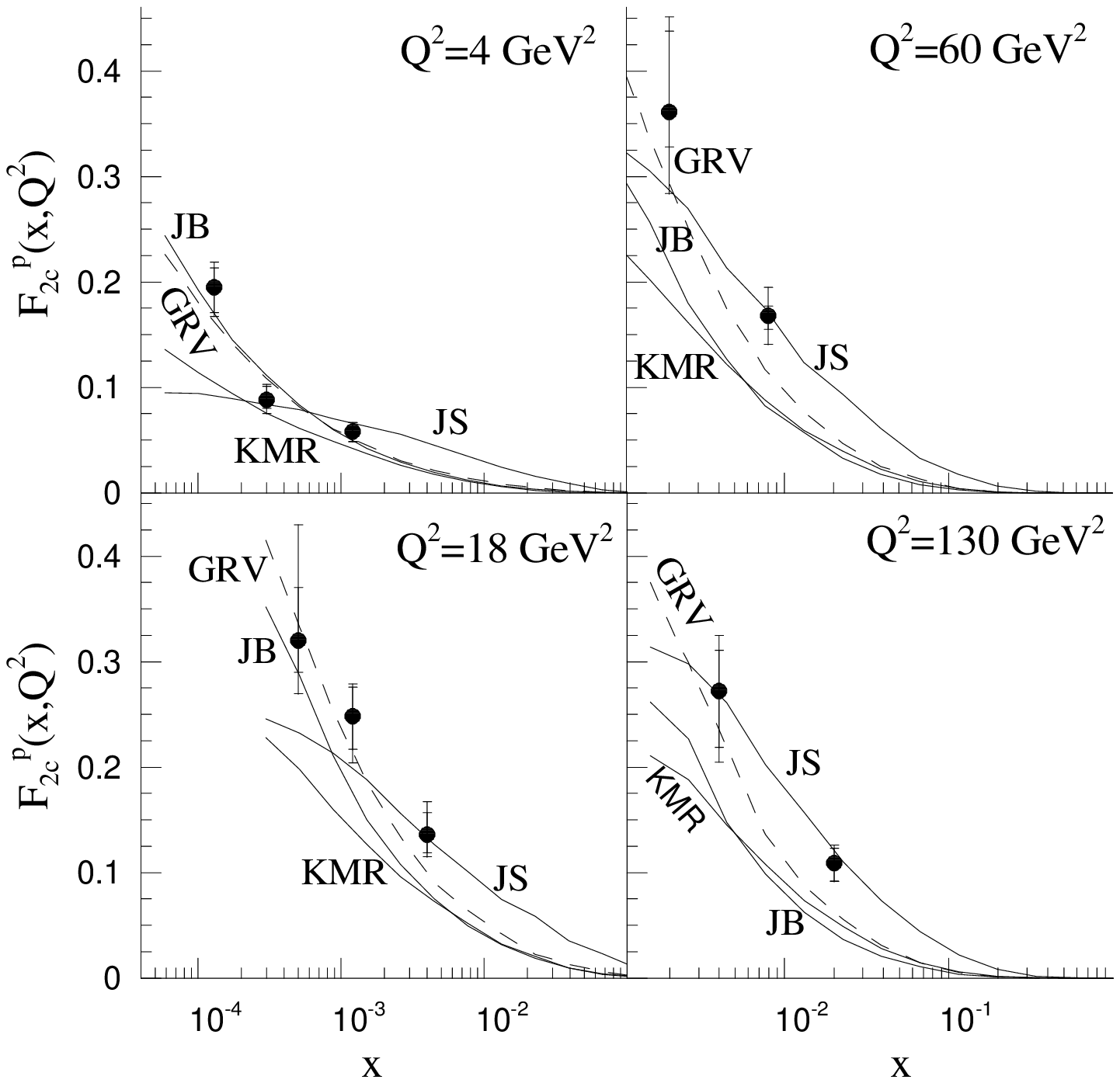}
\vspace*{12.0cm}
 \caption[*]{The SF $F_{2c}^p(x_B,Q^2)$ as a
function of $x_B$ at the $Q^2$=4, 18, 60 and 130 GeV$^2$ compared
to ZEUS data}
\end{center}
\end{figure}

\end{document}